\documentclass[aps,prl,reprint,superscriptaddress,showpacs,showkeys]{revtex4-1}

\usepackage{color}
\usepackage{comment}
\usepackage{slashbox}
\usepackage{amsmath}
\usepackage{graphicx}
\usepackage{color}
\usepackage{makecell}

\begin{document}

\title{Enhanced Superconducting Transition Temperature in Electroplated Rhenium}

\author{D.\ P.\ Pappas}\email{David.Pappas@NIST.gov}
\affiliation{National Institute of Standards and Technology, Boulder, Colorado 80305, USA}

\author{D.\ E.\ David}
\affiliation{CIRES, University of Colorado, Boulder, Colorado 80309, USA}

\author{R.\ E.\ Lake}
\affiliation{National Institute of Standards and Technology, Boulder, Colorado 80305, USA}

\author{M.\ Bal}
\affiliation{National Institute of Standards and Technology, Boulder, Colorado 80305, USA}
\affiliation{Department of Physics, University of Colorado, Boulder, Colorado 80309, USA}

\author{R.\ B.\ Goldfarb}
\affiliation{National Institute of Standards and Technology, Boulder, Colorado 80305, USA}

\author{D.\ A.\ Hite}
\affiliation{National Institute of Standards and Technology, Boulder, Colorado 80305, USA}

\author{E.\ Kim}
\affiliation{Department of Physics, University of Nevada Las Vegas, Las Vegas, Nevada 89154, USA}

\author{H.\ -S.\ Ku}
\affiliation{National Institute of Standards and Technology, Boulder, Colorado 80305, USA}
\affiliation{Department of Physics, University of Colorado, Boulder, Colorado 80309, USA}

\author{J.\ L.\ Long}
\affiliation{National Institute of Standards and Technology, Boulder, Colorado 80305, USA}
\affiliation{Department of Physics, University of Colorado, Boulder, Colorado 80309, USA}

\author{C.\ R.\ H.\ McRae}
\affiliation{National Institute of Standards and Technology, Boulder, Colorado 80305, USA}
\affiliation{Department of Physics, University of Colorado, Boulder, Colorado 80309, USA}

\author{L.\ D.\ Pappas}
\affiliation{CIRES, University of Colorado, Boulder, Colorado 80309, USA}

\author{A.\ Roshko}
\affiliation{National Institute of Standards and Technology, Boulder, Colorado 80305, USA}

\author{J.\ G.\ Wen}
\affiliation{Center for Nanoscale Materials, Argonne National Laboratory, Lemont, IL 60439, USA}

\author{B.\ L.\ T.\ Plourde}
\affiliation{Department of Physics, Syracuse University, Syracuse, New York 13244, USA}

\author{I.\ Arslan}
\affiliation{Center for Nanoscale Materials, Argonne National Laboratory, Lemont, IL 60439, USA}

\author{X.\ Wu}
\affiliation{National Institute of Standards and Technology, Boulder, Colorado 80305, USA}
\affiliation{Department of Physics, University of Colorado, Boulder, Colorado 80309, USA}

\date{\today}

\begin{abstract}
We show that electroplated Re films in multilayers with noble metals such as Cu, Au, and Pd have an enhanced  superconducting critical temperature relative to previous methods of preparing Re. The dc resistance and magnetic susceptibility indicate a critical temperature of approximately 6 K. Magnetic response as a function of field at 1.8 K demonstrates type-II superconductivity, with an upper critical field on the order of 2.5~T. Critical current densities greater than $10^7$~A/m$^2$ were measured above liquid-helium temperature. Low-loss at radio frequency was obtained below the critical temperature for multilayers deposited onto resonators made with Cu traces on commercial circuit boards. These electroplated superconducting films can be integrated into a wide range of standard components for low-temperature electronics.
\end{abstract}

\pacs{74.70.−b,74.70.Ad,74.62.Bf,74.78.−w,74.78.Fk,74.81.Bd }

\maketitle

Development of new superconducting materials with increased critical temperatures and practical integration into circuits and interconnects is important for many low temperature applications and experiments. Examples include high-speed, superconducting classical computers at liquid-helium temperatures, i.e., $T\sim$4~K, where the main goal is to improve energy efficiency \cite{ERSFQ,RQL,C3,HPC_SFQ, CryoKotsuboIEEE2017,CryoRadebaughJPhysCondMat2009} and also large-scale quantum information systems \cite{Devoret_Science13, Reilly_npj2015} at ultra-low temperature, $T<$30 mK. For the ultra-low temperature systems, i.e., work involving adiabatic demagnetization and dilution refrigerators, it important to have superconducting technology that is easily integrated into the many dc~bias and radio-frequency (rf)~drive lines. This will minimize dissipation to reduce heat load on the cryostats and help in preservation of quantum information as it propagates through the system.

To that end, there is a small set of superconductors that are widely used at 4 K and to bridge the gap to ultra-low temperatures. These include Nb and its binary and ternary alloys such as Nb-N, Nb-Ti and Nb-Ti-N. These materials are useful in terms of high critical temperature, can be used in bulk or deposited as thin films, and can be connected using ultrasonic-wirebonding techniques. However, they tend to be difficult to work with mechanically and have poor soldering properties due to strong oxidation. While work-arounds to these problems exist, they are not easily integrated into standard circuit fabrication that typically employs primarily Cu and Au electroplated parts.

Other superconductors, e.g., Al, Pb, In,  Sb, Re, high critical temperature ($T_c$) oxides, etc., are useful in many applications but have limitations due to low $T_c$, low melting point, difficulty making connections, or other problems. One element in this latter category, Re, stands out as a promising candidate. 
Re is a transition metal that is resistant to oxidation, with a melting temperature of 3186~$^\circ$C \cite{CRC}. It is used widely in various industrial and scientific applications such as strengthening materials and in high-temperature thermocouples. Crystalline Re is a type-I superconductor with $T_c(\rm Re)\approx1.7$~K \citep{MitoEnhancedTc2016}. Under strain Re changes to a type-II superconductor and $T_c(\rm Re)$ can be enhanced up to about 3~K \cite{AlexseevskiiJETP1966, DauntRe1952,MitoEnhancedTc2016}. This is similar to effects seen in Re doped with W and Os \cite{ChuDoping1971}. Moreover, Chu points out in Ref.  \citenum{ChuDoping1971} that there is a singularity close to the Fermi energy, $E_F$, that could drive anomalous behavior in the superconducting transition temperature through alloying, and it has recently been noted that Re-Mo alloys have  $T_c$ up to 13~K \cite{TestardiHiTc1971,SundarRePE2016}.
While both Re-Mo and epitaxial Re can be used for low-loss rf~resonators in qubit circuits, they have no specific advantage over other traditional material in that application\cite{EpiQubitsWeidesAPL2011, SingMoReResonators2014}. 

On the other hand, it has been known for a long time that Re can be electropated in aqueous solution that is compatible with other standard noble metal plating \cite{FinkElectroplatingRe1934, RheniumPlatingLeviPatent1952,SadanaEP-AuSurfCoat1989, GamburgElectroReRJE2015, ZhulikovElectroReRJE2016, ChangElectroReJES2015}. In this process the Re is deposited at the cathode in over-potential conditions, resulting in the concurrent production of hydrogen in the plating region. This brings up the topic of whether H can be incorporated into the Re and how that might affect its superconducting properties \cite{GorkovRevModPhys2018}. 

Here, we report on the superconducting nature of electroplated Re including electrical resistance, critical current, magnetization, and rf~loss. We prepare the Re in multilayers with noble metals because carbonaceous deposits tend to form on bare Re films. Cu, Au, and Pd were used in order to see if there were element-specific interface effects. Re films grown on both Au/Si and Cu-clad circuit-boards were studied. Type-II superconducting behavior is obtained with high critical-temperatures and critical-current densities and low rf~loss.   

The Re was deposited using a dc power supply in the constant-current mode at 8 A/dm$^2$ from an aqueous solution containing 11 g/L of KReO$_4$ with the pH adjusted to 0.9 using sulfuric acid, according to the recipe in Ref.   \citenum{FinkElectroplatingRe1934}.  Deposition time was used to control layer thickness. The anodes were platinized titanium. Moderate stirring was done with a magnetic bar coated with polytetrafluoroethylene (PTFE)  on a standard laboratory hot-plate stirrer at temperatures of 25-30~$^\circ$C.  Under these conditions the Re deposits were shiny and smooth, typically dark gray, with good adhesion. The various sample parameters are reported in Table \ref{TableSamples}. 

\begin{table}[h]
\begin{tabular}{ | p{2.9cm} | p{1cm} | p{3cm} | p{1cm} |  }
\hline
\backslashbox{Sample}{Layer}& Bottom & Middle  & Top\\ 
\hline

1 \makecell{Vacuum deposited\\ trilayer on Si } & 200 Au & 300 Re& 100 Au\\ 
\hline

2 \makecell{Electroplated\\ bilayer on Si }  & 200 Au& 300 Re &n/a\\ 
\hline

3 \makecell{Electroplated\\ trilayer on Si } &200 Au &300 Re &75 Au \\ 
\hline

4 \makecell{Electroplated \\multilayer on Si} &200 Au&(20 Au+75 Re)$\times$10& 75 Au \\ 
\hline

5 \makecell{Electroplated \\multilayer \\on Cu/PTFE} &200 Au &(20 Au+75 Re)$\times$10&75 Au \\ 
\hline

6 \makecell{Electroplated \\multilayer \\on Cu/PTFE} & N/A &(500 Cu+75 Re)$\times$5&500 Cu \\ 
\hline

7 \makecell{Electroplated \\multilayer \\on Cu/PTFE} & N/A &(500 Pd+75 Re)$\times$5&500 Pd \\ 
\hline

\end{tabular}
\caption{Sample composition and thicknesses (nm). For sample 1, a quartz crystal thickness monitor was used. For Samples 3 and 4, thicknesses were obtained directly from STEM. Nominal thicknesses are given for Samples 2, 5, 6 based on times derived from 3 and 4. The Au, Cu, and Pd were grown using standard electroplating solutions with the exceptions being the Au metallization with 5 nm Ti adhesion layer on the Si in Samples 2-4, and  the vacuum prepared trilayers for Sample 1.}
\label{TableSamples}
\end{table}

To begin the study, we first grew a reference, Sample 1, of Au/Re/Au/prime-Si in vacuum using e-beam evaporation for the Au and sputter-deposition for the Re. This sample was used to compare against standard Re preparation methods. 

Electroplated bilayers, trilayers, and  multilayers of Re and Au where then grown on vacuum-prepared Au/Si substrates, Samples 2-4. A 20 nm seed layer of electroplated Au was typically applied before the Re. 

Finally, multilayers of Re/Cu, Re/Au, and Re/Pd were electroplated directly onto commercial Cu/PTFE circuit boards, Samples 5-7. The boards were also patterned with resonators in order to test the rf~properties of the multilayers. PTFE was used for its low loss tangent, tan$\ \delta=6.8\times10^{-4}$, at 15~K  \cite{PTFE_LossMazierskafIEEE2005}. 

Of particular concern in many low-temperature measurements is the dc~resistance and rf~loss that can create ohmic heating and degradation of quantum information. Therefore, we characterized both types of transport properties. The measurements were conducted in both a liquid-He dewar-based system and an adiabatic demagnetization refrigerator (ADR). The samples on Si were measured with a dc 4-point probe. For the films on the 35~$\mu$m thick copper traces, it was necessary to use ac transport with a lock-in amplifier due to their low resistance. 

The resistivity of the sputtered-Re layers on Si was measured on $\sim$0.6 mm slices cleaved from 5 mm wide strips.  The $T_c$ taken to be the first inflection point in the R vs. T curves. Samples were contacted by either wirebonding (25 $\mu$m Al wire for low current measurements) or soldering (standard Cu wire for high current measurements). All samples could be soldered using 60:40 Sn-Pb. No problems were encountered wirebonding to the thick Cu whereas the for Pd- and Au-capped samples Al-wirebonds did not adhere as well. 

\begin{figure*}
\centering
\includegraphics[trim= {0 0 0 0}, clip, width = 12cm]{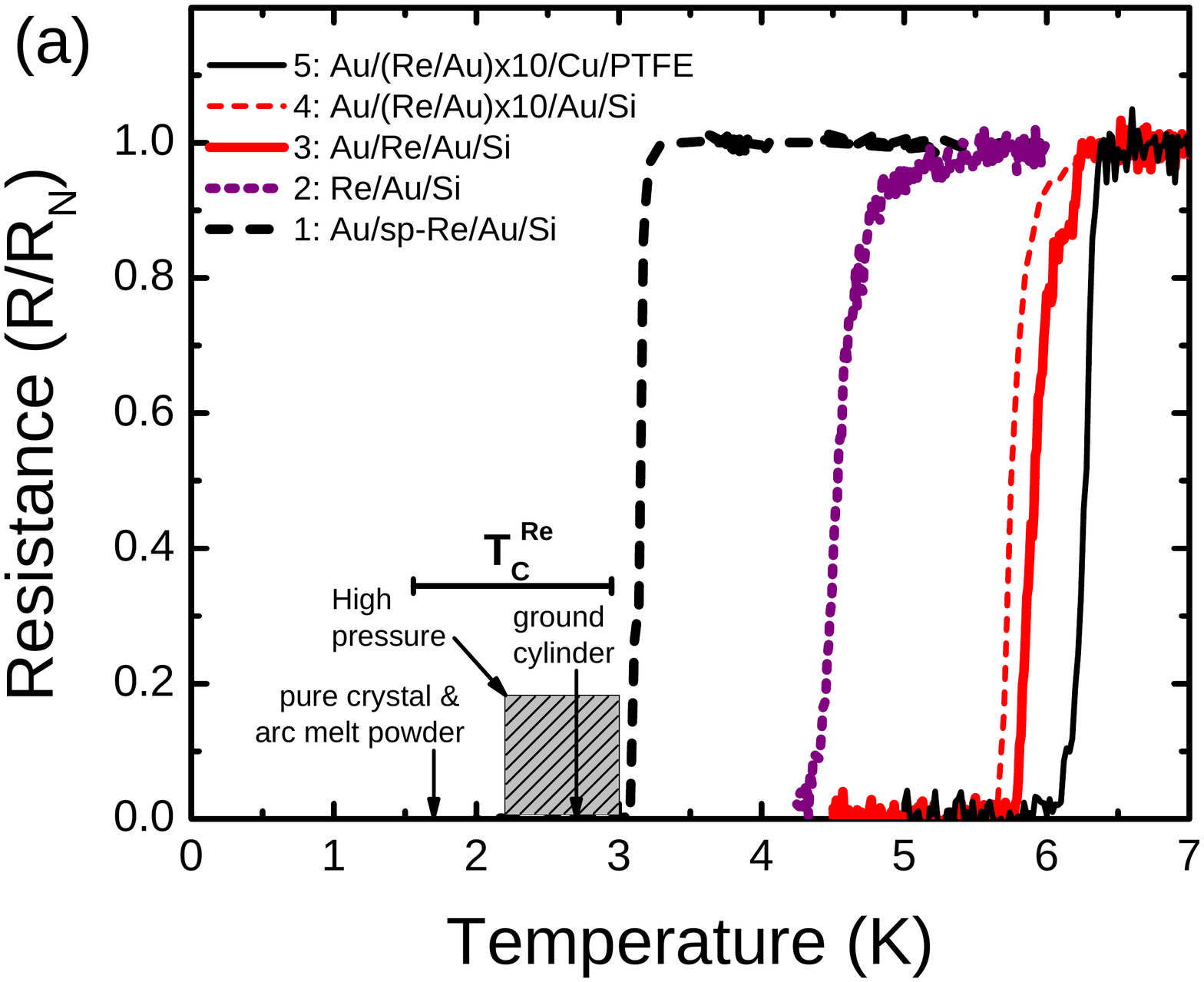}
\includegraphics[trim= {0 0 0 0}, clip, width = 12cm]{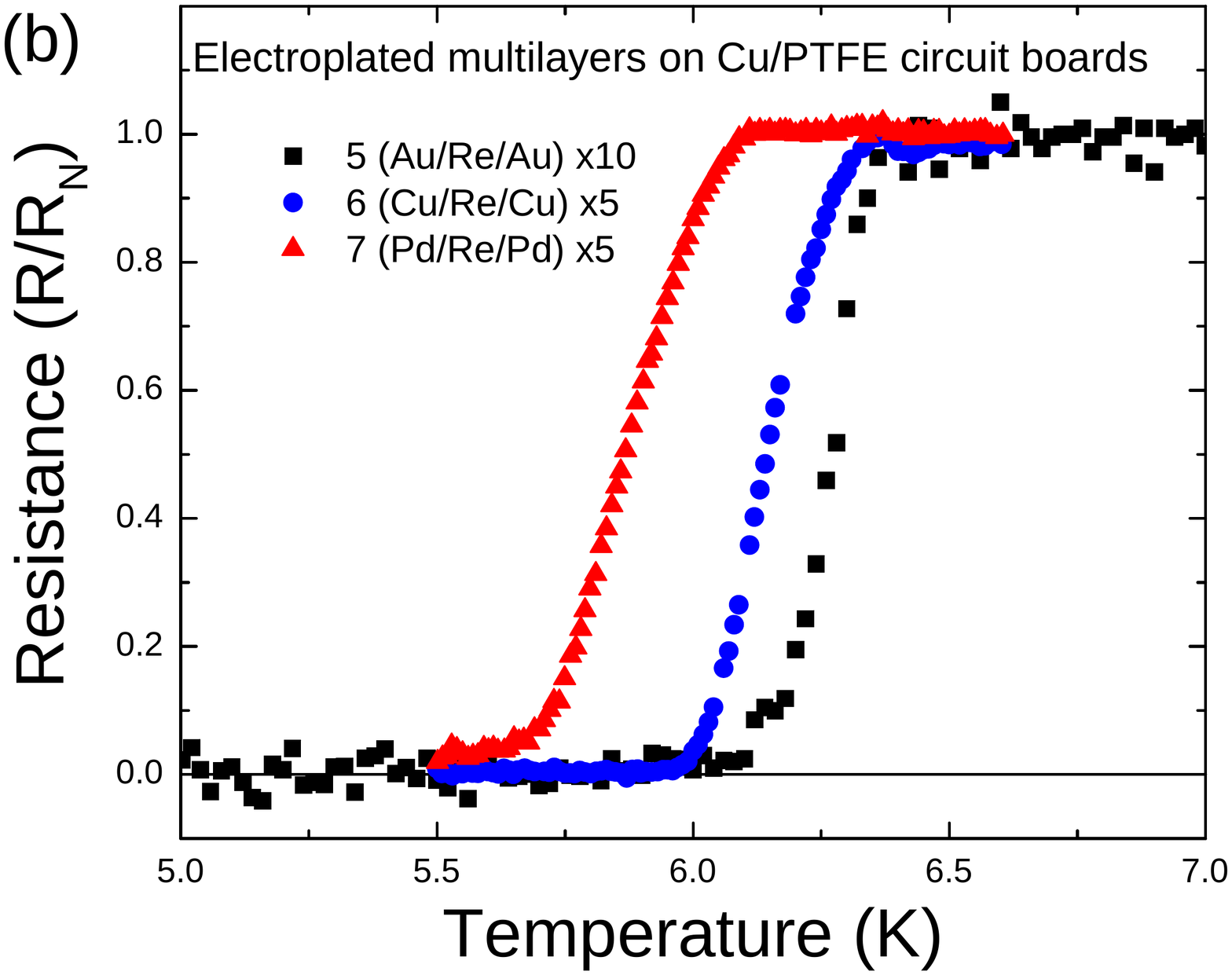}
\caption{Resistance vs. temperature. Panel (a) Data from Refs. \citenum {MitoEnhancedTc2016, ChuDoping1971} show $T_c$(Re) from the literature, curves show data from Au/Re samples described in Table \ref{TableSamples}; Panel (b) comparison of critical behavior for Re in Cu, Au, and Pd multilayers, Samples 5-7.}
\label{FigRvT}
\end{figure*}

Turning to the data, for the sputter-deposited Re samples we see a sharp transition to zero resistance in Fig. \ref{FigRvT}(a)1, with $T_c$ slightly exceeding that of the highly strained Re from Ref.  \citenum{MitoEnhancedTc2016}. This can be attributed to Au-Re interfacial strain that tends to expand the Re unit cell. Subsequent high temperature annealing, up to 400~$^\circ$C, of this sample showed that $T_c$ is stable within 0.1~K. This is in line with the high melting temperature of Re and its immiscibility with Au \cite{Okamoto1984, ShunkBinaryAlloys1985, Predel1991}. 

More surprisingly, for the electroplated samples on Si, a progression of $T_c$ up to approximately 6~K was obtained. First, with the Re/Au bilayer sample in  Fig. \ref{FigRvT}(a)2, we measured a range of transition temperatures from $T$=4.3~K to 4.7~K. These samples tend to tarnish over a period of a few weeks due to the exposed Re surface. Therefore, we moved to capping the samples with electroplated Au films. This resulted in more stable films and reproducible $T_c$ measurements. Moreover, $T_c$ in these layered samples increased to $T_c>5$~K, with multiple steps in the transition in some samples, as shown in Fig.  \ref{FigRvT}(a)3. Experiments on these films, including sputtering the top Au film off and using much thicker Au layers, tend to depress the $T_c$. Subsequent samples grown as multilayers on Si, Fig.  \ref{FigRvT}(a)4, demonstrated even higher reproducibility and enhancement of $T_c$. 

We then looked at the multilayered films on commercial circuit boards, Fig.  \ref{FigRvT}(a)5. Here, as $T$ decreases we see the first drop in the resistance for $T\approx6.1$~K. However, there are slight steps in these transitions with low resistance tails. The tails go down an additional 0.1 -- 0.5~K, depending on the current, before the transition to zero resistance.

In light of the changes of $T_c$(Re) due to strain, as outlined above, the properties of the Re grown with a series of metal films, i.e., Cu, Au, and Pd were compared. These metals all share a close-packed structure, fcc, that is similar to the hcp structure of Re. However, the nearest-neighbor distance in the close-packed plane for these three is 0.256, 0.275, and 0.288 nm, respectively \cite{Kittel}. This spans compressive, low, and expansive strain relative to the 0.274 nm near-neighbor spacing for elemental Re \cite{Kittel}. As shown in Fig. \ref{FigRvT}(b)1-3, all three types of multilayered films demonstrate enhancement of $T_c$. Both Re/Au and Re/Cu have a sharp drop in resistance well above 6~K, whereas the Re/Pd samples have a slightly lower, more rounded transition. This may be due to the fact that Re, Cu, and Au are immiscible, whereas RePd tends to alloy\cite{ShunkBinaryAlloys1985}. 

\begin{figure*}
\centering
\includegraphics[trim= {0 0 0 0}, clip, width = 15cm]{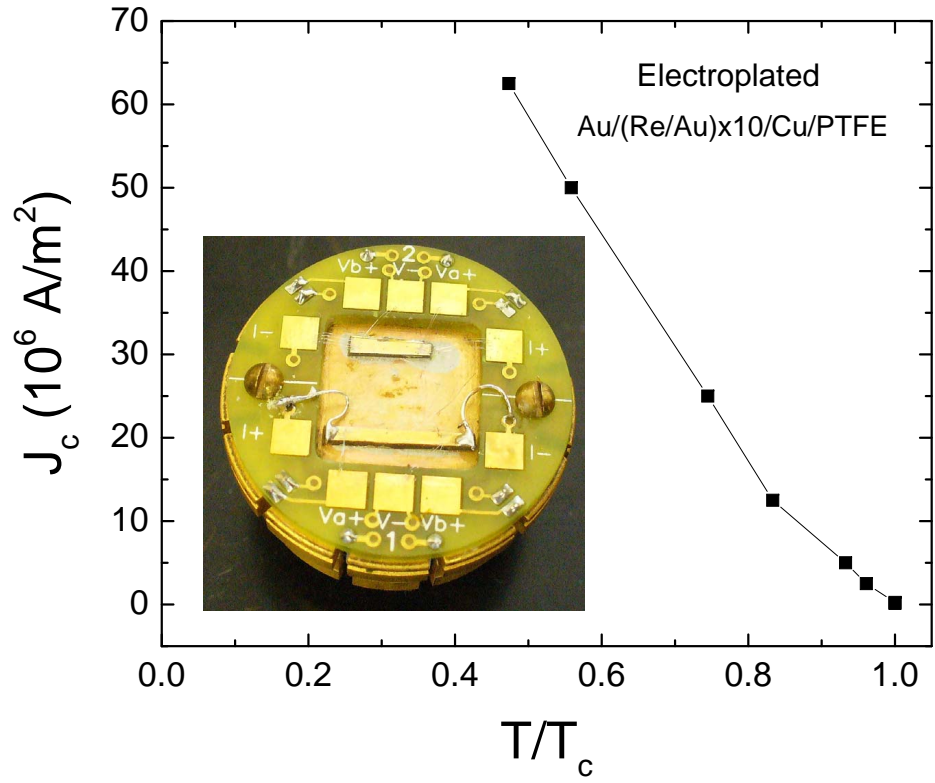}
\caption{Critical current $J_c$~vs.~ normalized temperature~$t=T/T_c$~for as-prepared electroplated multilayer, (Au/Re)x10/Au/PTFE, Sample 5. Inset shows two samples mounted; bottom sample with current leads soldered for high currents and voltage leads wirebonded; top sample is wirebonded on both current and voltage leads. 
}
\label{FigIcVt}
\end{figure*}

The versatility in connecting to the samples by soldering and wirebonding was important. This allowed us to achieve good thermalization while high current densities, $J$ are applied through the soldered contacts. In the same measurement,  voltage leads could be wirebonded, as shown in the inset of Fig. \ref{FigIcVt}. For samples on Si substrates we measured critical current $J_c \approx 2.5$ to $5 \times10^8$ A/m$^2$ for trilayers and multilayers, respectively. For films on Cu/PTFE, data in Fig. \ref{FigIcVt}, we saw lower $J_c \sim 10^7$ A/m$^2$ due to the low temperature tails. 

\begin{figure*}
\includegraphics[trim = {0cm 0 0 0},clip, width = 15cm]{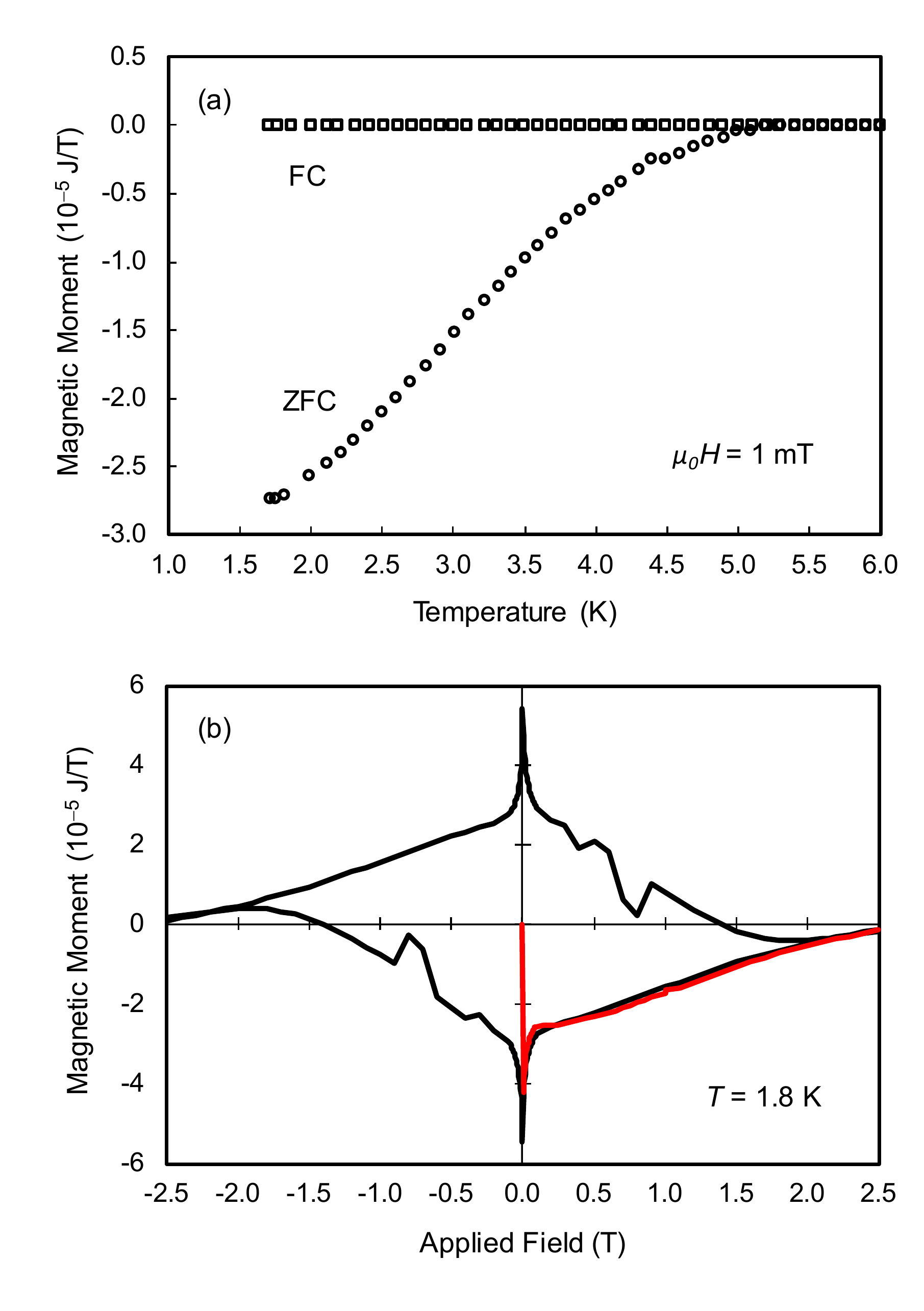} 
\caption{(a) Zero-field cooled~(ZFC) and field-cooled~(FC) magnetic moment as functions of increasing temperature measured in $\mu_0 H=1.0$~mT.  The critical temperature is $T_c=5.4$~K. (b) Magnetic hysteresis loop measured at $T=\ 1.8$~K. The initial curve with increasing field is in red. The upper critical field is $\mu_0 H_{c2}\approx2.5$~T. Flux penetration occurs below 0.4~mT. 
}
\label{FigMagnetization}
\end{figure*}

While it is significant that the resistance drops to zero, it does not unequivocally demonstrate that the films are superconducting over their entire area. Magnetic measurements, which access two dimensions, are able to distinguish this behavior. For this study, the Au/(AuRe)$\times$10/Si multilayer film, Sample 4, was cleaved into a rectangular sample, 5.0 mm $\times$ 5.6 mm, for measurement in a magnetometer based on a superconducting quantum interference device (SQUID). In magnetic measurements of superconductors, the apparent magnetic moment is due to the field produced by shielding currents and trapped magnetic flux. The moment was measured in a perpendicular field of $\mu_0H=1$~mT as a function of increasing temperature after cooling in zero field~(ZFC), 
and after cooling in a field~(FC) of $\mu_0H=1$~mT, Fig. \ref{FigMagnetization}(a). The critical temperature, $T_c=5.4$~K, is the end point of a broad transition from the shielding state to the normal state seen in the ZFC curve. The almost zero FC moment for $T<T_c$ is indicative of an incomplete Meissner effect (incomplete expulsion of magnetic flux upon cooling below $T_c$) due to strong flux pinning. 

A hysteresis loop of magnetic moment as a function of field was measured at $T=1.8$~K, Fig.  \ref{FigMagnetization}(b). The shape of the loop is characteristic of type-II superconductors. The magnetic moment approaches zero at an upper critical field $\mu_0 H_{c2}\approx2.5$~T. Symmetrical flux jumps are evident in the descending branches of the loop. A series of minor loops (not shown) with progressively higher maximum fields indicates the onset of hysteresis, and therefore flux penetration and pinning, below only 0.4~mT. This is likely due to the large demagnetizing factor for the film in perpendicular field \cite{Tinkham}.
 
\begin{figure*}
\centering
\includegraphics[trim= {0 0 0 0}, clip, width = 15cm]{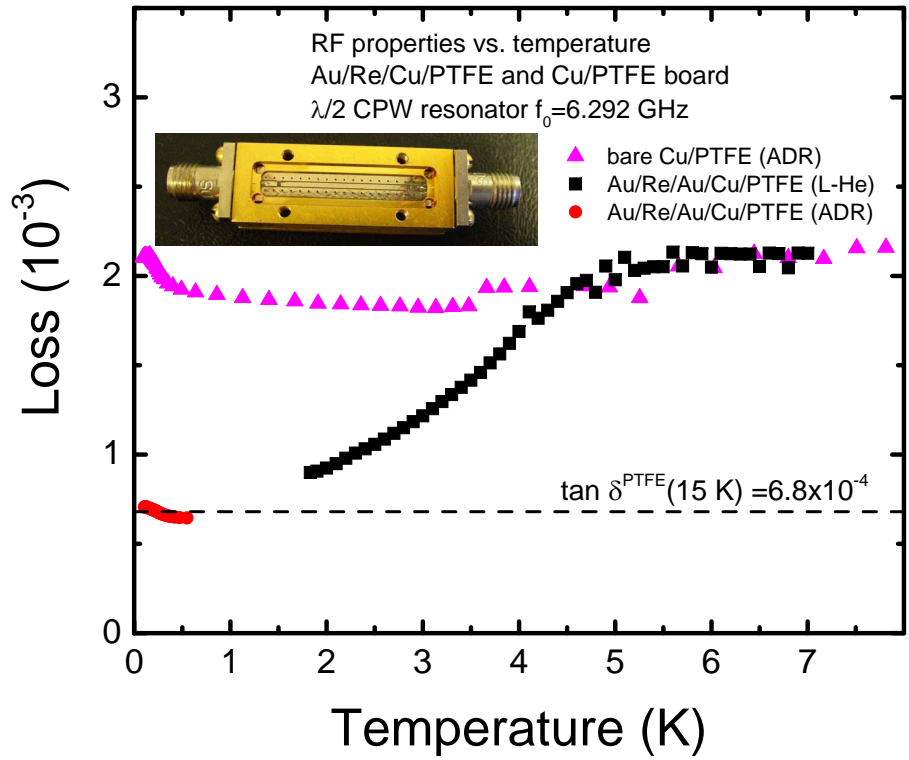}
\caption{Quality factor of Au/(Re/Au)x10/Cu/PTFE (Sample 5) compared to bare Cu/PTFE on a grounded coplanar resonator. The inset shows a Au/Re circuit board soldered into a Au-plated Cu box connectorized with SMA ports. }
\label{FigLvT}
\end{figure*}

For most superconducting applications, notably SQUID magnetometry and quantum computing, it is also important to establish that the films maintain low loss at rf into the gigahertz regime. Therefore, resonator loss tangent tan$\ \delta=1/{Q_T}$ vs. temperature was compared for Au/Re plated (Sample 5) vs.  bare copper traces. Here, $Q_T$ is the total quality factor of the resonator. The samples were placed in a magnetically shielded environment and cooled to low temperature. The rf measurements were made using a vector network analyzer. The $Q_T$ was determined using a a Lorenzian fit \cite{ProbstLorenztianRSI2015} to the resonance from a 1~mm wide, 18.2~mm long grounded-coplanar-waveguide resonator on a  0.59 mm thick PTFE board. The width and gap of the resonators were 0.81 and 0.076 mm, respectively. The coupling quality factor of the resonator to the feedline, $Q_C\sim2\times10^4$. Because the $Q_T$ was significantly lower than $Q_C$, the internal loss tangent tan$\ \delta_i=1/Q_i\approx1/Q_T$, where $Q_i$ is the internal quality factor. As shown in Fig. \ref{FigLvT}, for the bare copper board tan$\ \delta_i\approx2\times10^{-3}$  for $T<8$~K. Below about $T=250$ mK a slight increase appears. This increase is consistent with two-level system (TLS) loss \cite{TLS_PappasIEEE2011}, most likely in the PTFE. 

For the Re/Au electroplated circuit board, on the other hand, a decrease of tan$\ \delta_i$ for $T<5$~K is seen in Fig. \ref{FigLvT}. This is expected in the case of a superconducting transition in the Re-Au plating. The loss drops to that of PTFE at low temperature  \cite{PTFE_LossMazierskafIEEE2005}, and the TLS increase appears again at very low temperature. This shows that the loss is limited by the material of the circuit board rather than the metal traces. Therefore, development of new, low-loss board materials is now an important consideration for these circuits. On the other hand, these data show that plating of metallic components will significantly improve their performance at rf as well as dc.

In conclusion, this study shows that electroplated Re is superconducting with an enhanced $T_c\approx6$~K. This enhancement occurs when the Re is adjacent on both sides with noble metals, making it compatible with standard Cu and Au plated circuitry. It is weakly dependent on the type of metal, indicating that symmetry breaking at the interfaces likely plays a role along with possible bulk effects such as hydrogen incorporation, and nano-structure  \cite{Hasnain2016, GamburgElectroReRJE2015, GorkovRevModPhys2018}. This is corroborated by ongoing studies (not shown) of time-of-flight secondary-ion mass spectroscopy, which reveals high concentrations of H in the Re, and scanning transmission electron microscopy, which shows electron transmission through the Re layers with a striated pattern perpendicular to the plane of the film. Our own density-functional-theory calculations confirm the immiscibility of Au/Re and point to the possibility of an increase in the density of states near $E_F$  as the Re lattice expands, in agreement with Ref.   \citenum{ChuDoping1971}. These ongoing studies may help identify the relationship between morphology and superconductivity in Re. 

\section{Acknowledgments}
We had helpful discussions with Xingzhong Li, Tom Ohki, Jerry Chow, Gene Hilton, Kyle McKay, and Bahman Sarabi. We and thank Paul Blanchard and Chunsheng Tian at EAG Laboratories for preparing samples and measuring TOF-SIMS. This work was supported by the Intelligence Advanced Research Projects Activity (IARPA) LogiQ Program and the NIST Quantum Based Metrology Initiative. Use of the Center for Nanoscale Materials, an Office of Science user facility, was supported by the U.S. Department of Energy, Office of Science, Basic Energy Sciences, under Contract No. DE-AC02-06CH11357. This work is a contribution of NIST, not subject to copyright. 

\bibliography{ReElectroplated}

\begin{thebibliography}{33}
\expandafter\ifx\csname natexlab\endcsname\relax\def\natexlab#1{#1}\fi
\expandafter\ifx\csname bibnamefont\endcsname\relax
  \def\bibnamefont#1{#1}\fi
\expandafter\ifx\csname bibfnamefont\endcsname\relax
  \def\bibfnamefont#1{#1}\fi
\expandafter\ifx\csname citenamefont\endcsname\relax
  \def\citenamefont#1{#1}\fi
\expandafter\ifx\csname url\endcsname\relax
  \def\url#1{\texttt{#1}}\fi
\expandafter\ifx\csname urlprefix\endcsname\relax\def\urlprefix{URL }\fi
\providecommand{\bibinfo}[2]{#2}
\providecommand{\eprint}[2][]{\url{#2}}

\bibitem[{\citenamefont{Mukhanov}(2011)}]{ERSFQ}
\bibinfo{author}{\bibfnamefont{O.~A.} \bibnamefont{Mukhanov}},
  \bibinfo{journal}{IEEE Trans. Appl. Supercond.}
  \textbf{\bibinfo{volume}{21}}, \bibinfo{pages}{760} (\bibinfo{year}{2011}).

\bibitem[{\citenamefont{Herr et~al.}(2011)\citenamefont{Herr, Herr, Oberg, and
  Ioannidis}}]{RQL}
\bibinfo{author}{\bibfnamefont{Q.~P.} \bibnamefont{Herr}},
  \bibinfo{author}{\bibfnamefont{A.~Y.} \bibnamefont{Herr}},
  \bibinfo{author}{\bibfnamefont{O.~T.} \bibnamefont{Oberg}}, \bibnamefont{and}
  \bibinfo{author}{\bibfnamefont{A.~G.} \bibnamefont{Ioannidis}},
  \bibinfo{journal}{J. Appl. Phys.} \textbf{\bibinfo{volume}{109}},
  \bibinfo{pages}{Art. No. 103903} (\bibinfo{year}{2011}).

\bibitem[{\citenamefont{Manheimer}(2015)}]{C3}
\bibinfo{author}{\bibfnamefont{M.~A.} \bibnamefont{Manheimer}},
  \bibinfo{journal}{IEEE Trans. Appl. Supercond.}
  \textbf{\bibinfo{volume}{25}}, \bibinfo{pages}{Art. No. 1301704}
  (\bibinfo{year}{2015}).

\bibitem[{\citenamefont{Holmes et~al.}(2013)\citenamefont{Holmes, Ripple, and
  Manheimer}}]{HPC_SFQ}
\bibinfo{author}{\bibfnamefont{D.~S.} \bibnamefont{Holmes}},
  \bibinfo{author}{\bibfnamefont{A.~L.} \bibnamefont{Ripple}},
  \bibnamefont{and} \bibinfo{author}{\bibfnamefont{M.~A.}
  \bibnamefont{Manheimer}}, \bibinfo{journal}{IEEE Trans. Appl. Supercond.}
  \textbf{\bibinfo{volume}{23}}, \bibinfo{pages}{Art. No. 1701610}
  (\bibinfo{year}{2013}).

\bibitem[{\citenamefont{Kotsubo et~al.}(2017)\citenamefont{Kotsubo, Radebaugh,
  Hendershott, Bonczyski, Wilson, Nam, and Ullom}}]{CryoKotsuboIEEE2017}
\bibinfo{author}{\bibfnamefont{V.}~\bibnamefont{Kotsubo}},
  \bibinfo{author}{\bibfnamefont{R.}~\bibnamefont{Radebaugh}},
  \bibinfo{author}{\bibfnamefont{P.}~\bibnamefont{Hendershott}},
  \bibinfo{author}{\bibfnamefont{M.}~\bibnamefont{Bonczyski}},
  \bibinfo{author}{\bibfnamefont{B.}~\bibnamefont{Wilson}},
  \bibinfo{author}{\bibfnamefont{S.~W.} \bibnamefont{Nam}}, \bibnamefont{and}
  \bibinfo{author}{\bibfnamefont{J.~N.} \bibnamefont{Ullom}},
  \bibinfo{journal}{IEEE Trans. Appl. Supercond.}
  \textbf{\bibinfo{volume}{27}}, \bibinfo{pages}{Art. No 9500405}
  (\bibinfo{year}{2017}), ISSN \bibinfo{issn}{1051-8223}.

\bibitem[{\citenamefont{Radebaugh}(2009)}]{CryoRadebaughJPhysCondMat2009}
\bibinfo{author}{\bibfnamefont{R.}~\bibnamefont{Radebaugh}},
  \bibinfo{journal}{J. Phys.: Condens. Matter} \textbf{\bibinfo{volume}{21}},
  \bibinfo{pages}{Art. No. 164219} (\bibinfo{year}{2009}).

\bibitem[{\citenamefont{Devoret and Schoelkopf}(2013)}]{Devoret_Science13}
\bibinfo{author}{\bibfnamefont{M.~H.} \bibnamefont{Devoret}} \bibnamefont{and}
  \bibinfo{author}{\bibfnamefont{R.~J.} \bibnamefont{Schoelkopf}},
  \bibinfo{journal}{Science} \textbf{\bibinfo{volume}{339}},
  \bibinfo{pages}{1169} (\bibinfo{year}{2013}).

\bibitem[{\citenamefont{Reilly}(2015)}]{Reilly_npj2015}
\bibinfo{author}{\bibfnamefont{D.~J.} \bibnamefont{Reilly}},
  \bibinfo{journal}{npj Quantum Information} \textbf{\bibinfo{volume}{1}},
  \bibinfo{pages}{15011} (\bibinfo{year}{2015}).

\bibitem[{\citenamefont{Haynes}(2011)}]{CRC}
\bibinfo{author}{\bibfnamefont{W.~M.} \bibnamefont{Haynes}},
  \emph{\bibinfo{title}{CRC Handbook of Chemistry and Physics (92nd ed.)}}
  (\bibinfo{publisher}{CRC Press}, \bibinfo{address}{U.S.A.},
  \bibinfo{year}{2011}), ISBN \bibinfo{isbn}{ISBN 1439855110.}

\bibitem[{\citenamefont{Mito et~al.}(2016)\citenamefont{Mito, Matsui, Tsuruta,
  Yamaguchi, Nakamura, Deguchi, Shirakawa, Adachi, Yamasaki, Iwaoka
  et~al.}}]{MitoEnhancedTc2016}
\bibinfo{author}{\bibfnamefont{M.}~\bibnamefont{Mito}},
  \bibinfo{author}{\bibfnamefont{H.}~\bibnamefont{Matsui}},
  \bibinfo{author}{\bibfnamefont{K.}~\bibnamefont{Tsuruta}},
  \bibinfo{author}{\bibfnamefont{T.}~\bibnamefont{Yamaguchi}},
  \bibinfo{author}{\bibfnamefont{K.}~\bibnamefont{Nakamura}},
  \bibinfo{author}{\bibfnamefont{H.}~\bibnamefont{Deguchi}},
  \bibinfo{author}{\bibfnamefont{N.}~\bibnamefont{Shirakawa}},
  \bibinfo{author}{\bibfnamefont{H.}~\bibnamefont{Adachi}},
  \bibinfo{author}{\bibfnamefont{T.}~\bibnamefont{Yamasaki}},
  \bibinfo{author}{\bibfnamefont{H.}~\bibnamefont{Iwaoka}},
  \bibnamefont{et~al.}, \bibinfo{journal}{Sci. Rep.}
  \textbf{\bibinfo{volume}{6}}, \bibinfo{pages}{Art. No. 36337}
  (\bibinfo{year}{2016}).

\bibitem[{\citenamefont{Alekseevski\u{i}
  et~al.}(1967)\citenamefont{Alekseevski\u{i}, Mikheeva, and
  Tulina}}]{AlexseevskiiJETP1966}
\bibinfo{author}{\bibfnamefont{N.~E.} \bibnamefont{Alekseevski\u{i}}},
  \bibinfo{author}{\bibfnamefont{M.~N.} \bibnamefont{Mikheeva}},
  \bibnamefont{and} \bibinfo{author}{\bibfnamefont{N.~A.}
  \bibnamefont{Tulina}}, \bibinfo{journal}{Soviet Physics JETP}
  \textbf{\bibinfo{volume}{25}}, \bibinfo{pages}{575} (\bibinfo{year}{1967}).

\bibitem[{\citenamefont{Daunt and Smith}(1952)}]{DauntRe1952}
\bibinfo{author}{\bibfnamefont{J.~G.} \bibnamefont{Daunt}} \bibnamefont{and}
  \bibinfo{author}{\bibfnamefont{T.~S.} \bibnamefont{Smith}},
  \bibinfo{journal}{Phys. Rev} \textbf{\bibinfo{volume}{88}},
  \bibinfo{pages}{309} (\bibinfo{year}{1952}).

\bibitem[{\citenamefont{Chu et~al.}(1971)\citenamefont{Chu, McMillan, and
  Luo}}]{ChuDoping1971}
\bibinfo{author}{\bibfnamefont{C.~W.} \bibnamefont{Chu}},
  \bibinfo{author}{\bibfnamefont{W.~L.} \bibnamefont{McMillan}},
  \bibnamefont{and} \bibinfo{author}{\bibfnamefont{H.~L.} \bibnamefont{Luo}},
  \bibinfo{journal}{Phys. Rev. B} \textbf{\bibinfo{volume}{3}},
  \bibinfo{pages}{3757} (\bibinfo{year}{1971}).

\bibitem[{\citenamefont{Testardi et~al.}(1971)\citenamefont{Testardi, Hauser,
  and Read}}]{TestardiHiTc1971}
\bibinfo{author}{\bibfnamefont{L.~R.} \bibnamefont{Testardi}},
  \bibinfo{author}{\bibfnamefont{J.~J.} \bibnamefont{Hauser}},
  \bibnamefont{and} \bibinfo{author}{\bibfnamefont{M.~H.} \bibnamefont{Read}},
  \bibinfo{journal}{Solid State Comm.} \textbf{\bibinfo{volume}{9}},
  \bibinfo{pages}{1829} (\bibinfo{year}{1971}).

\bibitem[{\citenamefont{Sundar et~al.}(2016)\citenamefont{Sundar, Banik,
  Chandra, Chattopadhyay, Ganguli, Lodha, Pandey, Phase, and
  Roy}}]{SundarRePE2016}
\bibinfo{author}{\bibfnamefont{S.}~\bibnamefont{Sundar}},
  \bibinfo{author}{\bibfnamefont{S.}~\bibnamefont{Banik}},
  \bibinfo{author}{\bibfnamefont{L.~S.~S.} \bibnamefont{Chandra}},
  \bibinfo{author}{\bibfnamefont{M.~K.} \bibnamefont{Chattopadhyay}},
  \bibinfo{author}{\bibfnamefont{T.}~\bibnamefont{Ganguli}},
  \bibinfo{author}{\bibfnamefont{G.~S.} \bibnamefont{Lodha}},
  \bibinfo{author}{\bibfnamefont{S.~K.} \bibnamefont{Pandey}},
  \bibinfo{author}{\bibfnamefont{D.~M.} \bibnamefont{Phase}}, \bibnamefont{and}
  \bibinfo{author}{\bibfnamefont{S.~B.} \bibnamefont{Roy}},
  \bibinfo{journal}{J. Phys.: Condens. Matter} \textbf{\bibinfo{volume}{28}},
  \bibinfo{pages}{Art. No. 315502} (\bibinfo{year}{2016}).

\bibitem[{\citenamefont{Weides et~al.}(2011)\citenamefont{Weides, Kline,
  Vissers, Sandberg, Wisbey, Johnson, Ohki, and
  Pappas}}]{EpiQubitsWeidesAPL2011}
\bibinfo{author}{\bibfnamefont{M.~P.} \bibnamefont{Weides}},
  \bibinfo{author}{\bibfnamefont{J.~S.} \bibnamefont{Kline}},
  \bibinfo{author}{\bibfnamefont{M.~R.} \bibnamefont{Vissers}},
  \bibinfo{author}{\bibfnamefont{M.~O.} \bibnamefont{Sandberg}},
  \bibinfo{author}{\bibfnamefont{D.~S.} \bibnamefont{Wisbey}},
  \bibinfo{author}{\bibfnamefont{B.~R.} \bibnamefont{Johnson}},
  \bibinfo{author}{\bibfnamefont{T.~A.} \bibnamefont{Ohki}}, \bibnamefont{and}
  \bibinfo{author}{\bibfnamefont{D.~P.} \bibnamefont{Pappas}},
  \bibinfo{journal}{Appl. Phys. Lett.} \textbf{\bibinfo{volume}{99}},
  \bibinfo{pages}{Art. No. 262502} (\bibinfo{year}{2011}).

\bibitem[{\citenamefont{Singh et~al.}(2014)\citenamefont{Singh, Schneider,
  Bosman, Merkx, and Steele}}]{SingMoReResonators2014}
\bibinfo{author}{\bibfnamefont{V.}~\bibnamefont{Singh}},
  \bibinfo{author}{\bibfnamefont{B.~H.} \bibnamefont{Schneider}},
  \bibinfo{author}{\bibfnamefont{S.~J.} \bibnamefont{Bosman}},
  \bibinfo{author}{\bibfnamefont{E.~P.~J.} \bibnamefont{Merkx}},
  \bibnamefont{and} \bibinfo{author}{\bibfnamefont{G.~A.}
  \bibnamefont{Steele}}, \bibinfo{journal}{Appl. Phys. Lett.}
  \textbf{\bibinfo{volume}{105}}, \bibinfo{pages}{Art. No. 222601}
  (\bibinfo{year}{2014}).

\bibitem[{\citenamefont{Fink and Deren}(1934)}]{FinkElectroplatingRe1934}
\bibinfo{author}{\bibfnamefont{C.~G.} \bibnamefont{Fink}} \bibnamefont{and}
  \bibinfo{author}{\bibfnamefont{P.}~\bibnamefont{Deren}},
  \bibinfo{journal}{Trans. Electrochem. Soc.} \textbf{\bibinfo{volume}{66}},
  \bibinfo{pages}{471} (\bibinfo{year}{1934}).

\bibitem[{\citenamefont{Levi}(1952)}]{RheniumPlatingLeviPatent1952}
\bibinfo{author}{\bibfnamefont{R.}~\bibnamefont{Levi}},
  \emph{\bibinfo{title}{Rhenium plating}} (\bibinfo{year}{1952}),
  \bibinfo{note}{\uppercase{U.S.} Patent 2,616,840},
  \urlprefix\url{https://patentimages.storage.googleapis.com/8c/58/83/435234120f94b3/US2616840.pdf}.

\bibitem[{\citenamefont{Sadana and Wang}(1989)}]{SadanaEP-AuSurfCoat1989}
\bibinfo{author}{\bibfnamefont{Y.~N.} \bibnamefont{Sadana}} \bibnamefont{and}
  \bibinfo{author}{\bibfnamefont{Z.~Z.} \bibnamefont{Wang}},
  \textbf{\bibinfo{volume}{37}}, \bibinfo{pages}{419} (\bibinfo{year}{1989}).

\bibitem[{\citenamefont{Gamburg and Puryaeva}(2015)}]{GamburgElectroReRJE2015}
\bibinfo{author}{\bibfnamefont{A.~B.} \bibnamefont{Gamburg}, \bibfnamefont{Yu.
  D. and.~Drovosekovz}} \bibnamefont{and} \bibinfo{author}{\bibfnamefont{T.~P.}
  \bibnamefont{Puryaeva}}, \bibinfo{journal}{Russian Journal of
  Electrochemistry} \textbf{\bibinfo{volume}{51}}, \bibinfo{pages}{376}
  (\bibinfo{year}{2015}).

\bibitem[{\citenamefont{Zhulikov and Gamburg}(2016)}]{ZhulikovElectroReRJE2016}
\bibinfo{author}{\bibfnamefont{V.~V.} \bibnamefont{Zhulikov}} \bibnamefont{and}
  \bibinfo{author}{\bibfnamefont{Y.~D.} \bibnamefont{Gamburg}},
  \bibinfo{journal}{Russian Journal of Electrochemistry}
  \textbf{\bibinfo{volume}{52}}, \bibinfo{pages}{847} (\bibinfo{year}{2016}).

\bibitem[{\citenamefont{Chang et~al.}(2015)\citenamefont{Chang, Liang, Kao, and
  Lin}}]{ChangElectroReJES2015}
\bibinfo{author}{\bibfnamefont{S.-Y.} \bibnamefont{Chang}},
  \bibinfo{author}{\bibfnamefont{L.-P.} \bibnamefont{Liang}},
  \bibinfo{author}{\bibfnamefont{L.-C.} \bibnamefont{Kao}}, \bibnamefont{and}
  \bibinfo{author}{\bibfnamefont{C.-F.} \bibnamefont{Lin}},
  \bibinfo{journal}{Journal of The Electrochemical Society}
  \textbf{\bibinfo{volume}{162}}, \bibinfo{pages}{D96} (\bibinfo{year}{2015}).

\bibitem[{\citenamefont{Gor'kov}(2018)}]{GorkovRevModPhys2018}
\bibinfo{author}{\bibfnamefont{L.~P.} \bibnamefont{Gor'kov}},
  \bibinfo{journal}{Rev. Mod. Phys.} \textbf{\bibinfo{volume}{90}},
  \bibinfo{pages}{Art. No. 011001} (\bibinfo{year}{2018}).

\bibitem[{\citenamefont{Mazierska et~al.}(2005)\citenamefont{Mazierska, Jacob,
  Ledenyov, and Krupk}}]{PTFE_LossMazierskafIEEE2005}
\bibinfo{author}{\bibfnamefont{J.}~\bibnamefont{Mazierska}},
  \bibinfo{author}{\bibfnamefont{M.~V.} \bibnamefont{Jacob}},
  \bibinfo{author}{\bibfnamefont{D.}~\bibnamefont{Ledenyov}}, \bibnamefont{and}
  \bibinfo{author}{\bibfnamefont{J.}~\bibnamefont{Krupk}}, in
  \emph{\bibinfo{booktitle}{2005 Asia-Pacific Microwave Conference
  Proceedings}} (\bibinfo{year}{2005}), vol.~\bibinfo{volume}{4}, p.
  \bibinfo{pages}{2370}.

\bibitem[{\citenamefont{Okamoto and Massalski}(1984)}]{Okamoto1984}
\bibinfo{author}{\bibfnamefont{H.}~\bibnamefont{Okamoto}} \bibnamefont{and}
  \bibinfo{author}{\bibfnamefont{T.~B.} \bibnamefont{Massalski}},
  \bibinfo{journal}{Bulletin of Alloy Phase Diagrams}
  \textbf{\bibinfo{volume}{5}}, \bibinfo{pages}{383} (\bibinfo{year}{1984}),
  ISSN \bibinfo{issn}{0197-0216},
  \urlprefix\url{https://doi.org/10.1007/BF02872959}.

\bibitem[{\citenamefont{Shunk}(1985)}]{ShunkBinaryAlloys1985}
\bibinfo{author}{\bibfnamefont{F.~A.} \bibnamefont{Shunk}},
  \emph{\bibinfo{title}{Constitution of Binary Alloys, Second Supplement}}
  (\bibinfo{publisher}{McGraw-Hill Book Company}, \bibinfo{address}{U.S.A.},
  \bibinfo{year}{1985}), ISBN \bibinfo{isbn}{0-931690-20-X}.

\bibitem[{\citenamefont{Predel}(1991)}]{Predel1991}
\bibinfo{author}{\bibfnamefont{B.}~\bibnamefont{Predel}},
  \emph{\bibinfo{title}{Au-Re (Gold-Rhenium)}} (\bibinfo{publisher}{Springer
  Berlin Heidelberg}, \bibinfo{address}{Berlin, Heidelberg},
  \bibinfo{year}{1991}), p.~\bibinfo{pages}{1}, ISBN
  \bibinfo{isbn}{978-3-540-39444-0},
  \urlprefix\url{https://doi.org/10.1007/10000866_311}.

\bibitem[{\citenamefont{Kittel}(1976)}]{Kittel}
\bibinfo{author}{\bibfnamefont{C.}~\bibnamefont{Kittel}},
  \emph{\bibinfo{title}{Introduction to Solid State Physics, 5th edition}}
  (\bibinfo{publisher}{John Wiley \& Sons, Inc.}, \bibinfo{address}{U.S.A.},
  \bibinfo{year}{1976}), ISBN \bibinfo{isbn}{ISBN 0-471-49024-5}.

\bibitem[{\citenamefont{Tinkham}(1996)}]{Tinkham}
\bibinfo{author}{\bibfnamefont{M.}~\bibnamefont{Tinkham}},
  \emph{\bibinfo{title}{Introduction to Superconductivity, Second Edition}}
  (\bibinfo{publisher}{Dover Publications}, \bibinfo{address}{Mineola, New
  York}, \bibinfo{year}{1996}).

\bibitem[{\citenamefont{Probst et~al.}(2015)\citenamefont{Probst, Song, Bushev,
  Ustinov, and Weides}}]{ProbstLorenztianRSI2015}
\bibinfo{author}{\bibfnamefont{S.}~\bibnamefont{Probst}},
  \bibinfo{author}{\bibfnamefont{F.~B.} \bibnamefont{Song}},
  \bibinfo{author}{\bibfnamefont{P.~A.} \bibnamefont{Bushev}},
  \bibinfo{author}{\bibfnamefont{A.~V.} \bibnamefont{Ustinov}},
  \bibnamefont{and} \bibinfo{author}{\bibfnamefont{M.}~\bibnamefont{Weides}},
  \bibinfo{journal}{Rev. of Sci. Instrum.} \textbf{\bibinfo{volume}{86}},
  \bibinfo{pages}{024706} (\bibinfo{year}{2015}), ISSN
  \bibinfo{issn}{0197-0216}.

\bibitem[{\citenamefont{Pappas et~al.}(2011)\citenamefont{Pappas, Vissers,
  Wisbey, Kline, and Gao}}]{TLS_PappasIEEE2011}
\bibinfo{author}{\bibfnamefont{D.~P.} \bibnamefont{Pappas}},
  \bibinfo{author}{\bibfnamefont{M.~R.} \bibnamefont{Vissers}},
  \bibinfo{author}{\bibfnamefont{D.~S.} \bibnamefont{Wisbey}},
  \bibinfo{author}{\bibfnamefont{J.~S.} \bibnamefont{Kline}}, \bibnamefont{and}
  \bibinfo{author}{\bibfnamefont{J.}~\bibnamefont{Gao}}, \bibinfo{journal}{IEEE
  Trans. Appl. Supercond.} \textbf{\bibinfo{volume}{21}}, \bibinfo{pages}{871}
  (\bibinfo{year}{2011}).

\bibitem[{\citenamefont{Syed et~al.}(2016)\citenamefont{Syed, Webb, and
  Gray}}]{Hasnain2016}
\bibinfo{author}{\bibfnamefont{H.~M.} \bibnamefont{Syed}},
  \bibinfo{author}{\bibfnamefont{C.}~\bibnamefont{Webb}}, \bibnamefont{and}
  \bibinfo{author}{\bibfnamefont{E.~M.} \bibnamefont{Gray}},
  \bibinfo{journal}{Progress in Solid State Chemistry}
  \textbf{\bibinfo{volume}{44}}, \bibinfo{pages}{20} (\bibinfo{year}{2016}).

\end{thebibliography}
\bibliographystyle{apsrev}

\pagebreak

\pagebreak

\pagebreak

\pagebreak

\end{document}